\documentstyle[prl,aps,epsfig]{revtex}         % 2 col mode

\begin{document}
% define commands for international characters
\catcode`\ä = \active \catcode`\ö = \active \catcode`\ü = \active
\catcode`\Ä = \active \catcode`\Ö = \active \catcode`\Ü = \active
\catcode`\ß = \active \catcode`\é = \active \catcode`\è = \active
\catcode`\ë = \active \catcode`\ô = \active \catcode`\ê = \active
\catcode`\ø = \active \catcode`\ò = \active \catcode`\í = \active
\defä{\"a} \defö{\"o} \defü{\"u} \defÄ{\"A} \defÖ{\"O} \defÜ{\"U} \defß{\ss} \defé{\'{e}}
\defè{\`{e}} \defë{\"{e}} \defô{\^{o}} \defê{\^{e}} \defø{\o} \defò{\`{o}} \defí{\'{i}}
\draft               % preprint mode
\newcommand{\ifm}{interferometer }
\newcommand{\stz}{$|0\,\hbar k\rangle$ }
\newcommand{\stt}{$|2\,\hbar k \rangle$ }
\newcommand{\stmt}{$|-2\,\hbar k \rangle$ }
\newcommand{\stf}{$|4\,\hbar k \rangle$ }
\newcommand{\stmf}{$|-4\,\hbar k \rangle$ }
\newcommand{\om}{\omega_{\rm rec}}
% 2 col mode:
\twocolumn[\hsize\textwidth\columnwidth\hsize\csname
@twocolumnfalse\endcsname %\vspace{-5mm}

\title{Contrast Interferometry using Bose-Einstein Condensates to Measure $h/m$ and $\alpha$} \vspace{-5mm}
\author{S. Gupta, K. Dieckmann, Z. Hadzibabic and D. E. Pritchard}
\address{Department of Physics, MIT-Harvard Center for Ultracold
Atoms, and Research Laboratory
of Electronics, \\
MIT, Cambridge, MA 02139}
\date{\today}
\maketitle

\begin{abstract}
The kinetic energy of an atom recoiling due to absorption of a
photon was measured as a frequency using an interferometric
technique called ``contrast interferometry''. Optical standing
wave pulses were used as atom-optical elements to create a
symmetric three-path \ifm with a Bose-Einstein condensate. The
recoil phase accumulated in different paths was measured using a
single-shot detection technique. The scheme allows for additional
photon recoils within the \ifm and its symmetry suppresses several
random and systematic errors including those from vibrations and
ac Stark shifts. We have measured the photon recoil frequency of
sodium to $7\,$ppm precision, using a simple realization of this
scheme. Plausible extensions should yield a sufficient precision
to bring within reach a ppb-level determination of $h/m$ and the
fine structure constant $\alpha$.
\end{abstract}
\vskip1pc]

\narrowtext

%--- general introduction--dave--------------------------------------------
Comparison of accurate measurements of the fine structure constant
$\alpha$ in different subfields of physics - e.g. atomic physics,
QED and condensed matter physics - offer one of the few checks for
global errors across these different subfields. The $(g\!-\!2)$
measurement for the electron and positron together with QED
calculations, provides a $4\,$ppb \cite{dyck87,kino95} measurement
of $\alpha$. This has stood as the best measurement of the fine
structure constant since 1987. The second most accurate published
value of $\alpha$, at $24\,$ppb comes from condensed matter
experiments \cite{cage89}. This is worse by a factor of six,
limiting the scientific value of cross-field comparisons. A new
and more robust route based on atomic physics measurements has
emerged in the last decade \cite{tayl94}:

\begin{equation}\label{alpha}
  \alpha^2 = \left(\frac{e^2}{\hbar\,c}\right)^2 = \frac{2R_\infty}{c}\,\frac{h}{m_e} =
  \frac{2R_\infty}{c} \,\frac{M}{M_e} \,\frac{h}{m}\hspace{1mm}.
\end{equation}

The Rydberg constant $R_\infty$ is known to $0.008\,$ppb
\cite{schw99,udem97} and the electron mass $M_e$ to $0.7\,$ppb
\cite{beie02}. $M$ and $m$ are the mass of some test particle in
atomic and SI units respectively. Eq.$\,$\ref{alpha} offers the
possibility of a ppb level measurement of $\alpha$ if $M$ and
$h/m$ can be determined accurately.

$h/m$ can be measured by comparing the deBroglie wavelength and
velocity of a particle, as demonstrated by Kr\"{u}ger, whose
measurement using neutrons has yielded a value for $h/m_n$
accurate at $73\,$ppb \cite{krug98}. For an atom, $h/m$ can be
extracted from a measurement of the photon recoil frequency
\begin{equation}\label{omega}
  \om = {1\over 2}{\hbar\over m}k^2\hspace{1mm},
\end{equation}
where $k$ is the wavevector of the photon absorbed by the atom.
Recent experiments allow Eqs.$\,$\ref{alpha},$\,$\ref{omega} to be
applied to cesium. $M_{\rm Cs}$ is known to
$0.17\,$ppb\cite{brad99} and $k_{\rm Cs}$ to
$0.12\,$ppb\cite{udem99}. $\omega_{\rm rec,Cs}$ has been measured
at Stanford using an atom \ifm based on laser-cooled atoms
\cite{weis93,youn97,hens01} to $\approx 10\,$ppb \cite{foot1}.
Similar experiments are also possible with alkali atoms like
sodium and rubidium where $M$ has been measured \cite{brad99} and
$k$ is accurately accessible \cite{udem99}.
%--- end of the general introduction------------------------------------

%--- introduction of our interferometer---------------------------------
In this Letter, we demonstrate a new atom \ifm scheme which shows
promise for a high precision measurement of $\om$. Our symmetric
three-path configuration encodes the photon recoil phase in the
{\it contrast} of the interference fringes, rather than in their
{\it phase}. Because it is insensitive to the fringe phase, the
method is not sensitive to vibrations, accelerations or rotations.
The symmetry also suppresses errors from magnetic field gradients
and our use of only one internal state suppresses errors arising
from differences in the ac Stark shifts between different internal
states. A crucial aspect of this new \ifm is the use of atomic
samples with sub-recoil momentum distribution. We use a
Bose-Einstein condensate (BEC) as a bright atom source. This
allows the contrast oscillations to persist for many cycles,
permitting precise determination of the recoil phase in a single
``shot'' and also allows for extra photon recoils to be added
within the interferometer, increasing the recoil phase shift
quadratically.

The Stanford scheme to measure $\om$ uses different internal
states to separately address different \ifm paths, allowing a
linear increase of measurement precision by additional photon
recoils. However, vibrations and ac Stark shifts have been of
great concern in this scheme \cite{youn97}. An alternative \ifm to
measure the photon recoil using laser cooled atoms in a single
internal state was demonstrated using rubidium atoms
\cite{cahn97}. Like ours, this scheme also incorporates a
symmetric arrangement and operates by measuring contrast. This
\ifm should also suppress vibration noise and systematics arising
from ac Stark shifts between different internal states. However,
different paths cannot be individually addressed in this scheme,
making it difficult to extend to competitive precision. Our \ifm
extends these previous schemes and combines their advantages. The
precision of the Stanford scheme increases linearly with
additional recoils. Quadratic scaling schemes have been proposed
\cite{berm97} and demonstrated in a multi-path \ifm based on dark
states \cite{weit97}. However, the number of additional recoils in
this scheme, is limited by the internal atomic structure.

Our scheme is based on the asymmetric \ifm of
Fig.$\,$\ref{fig:scheme}(a). At time $t\!=\!0$ a BEC is split
coherently into two momentum components, \stt and $|0\,\hbar
k\rangle$, by a first order Bragg $\pi\!/2$-pulse \cite{gupt01}.
These are shown as paths 1 and 2 in the figure. At time $t\!=\!T$,
a second order Bragg $\pi$-pulse reverses the direction of path 1,
while leaving path 2 unaffected. Around $t\!=\!2\,T$, a moving
matter wave grating, with spatial periodicity $\lambda/2$
(wavevector $2k=\frac{2\pi}{\lambda/2}$), is formed due to the
overlap and interference of the two paths. The phase of this
grating at $2T$ is determined by the relative phase
$\Phi_1-\Phi_2=8\,\om\,T$, accumulated between paths 1 and 2 due
to the difference in their kinetic energies. A measurement of this
phase for different values of $T$ will then determine $\om$. If
the momentum of path 1 is increased $N$ times by additional photon
recoils, the corresponding grating phase will be
$\Phi_1-\Phi_2=N^28\,\om\,T$, leading to an $N^2$-fold improvement
in the measurement precision. The fringes of all atoms will be in
phase at $2T$, forming a high-contrast matter wave grating. This
dephases in a time $\frac{1}{k\Delta v}$, the coherence time,
where $\Delta v$ is the atomic velocity spread.
\begin{figure}[htbf]
\begin{center}
\vskip0mm \epsfxsize=80mm {\epsfbox{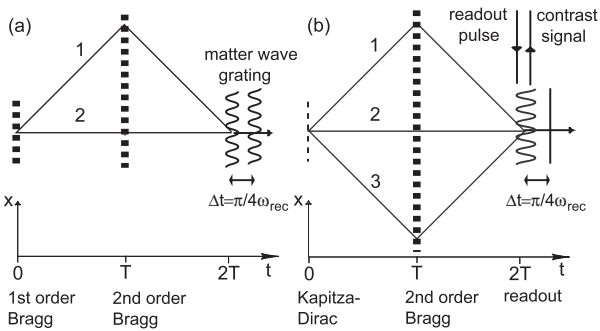}} \vskip1mm
\end{center}
\caption{Space-Time representation of the contrast interferometer.
(a) shows a simple 2-path \ifm sensitive to the photon recoil
phase. The $2k$ matter wave grating is shown at $2T$ and at
$2T+\pi/4\,\om$. The extension to the 3-path geometry is shown in
(b). The overall $2k$ grating has large contrast at $2T$ and zero
contrast at $2T+\pi/4\,\om$.} \label{fig:scheme}
\end{figure}
\vskip-3mm

Extension of this \ifm to a symmetric three-path arrangement is
shown in Fig.$\,$\ref{fig:scheme}(b). Three momentum states (paths
1, 2 and 3) are generated by replacing the first Bragg pulse with
a short Kapitza-Dirac pulse \cite{gupt01}. At $t\!=\!2T$, there
are now two matter wave gratings with period $\lambda/2$, one from
paths 1 and 2 and one from paths 2 and 3. These move in opposite
directions at a relative speed $4\hbar k/m$. If the maxima of the
two gratings line up to produce large contrast at time $t$, the
maxima of one will line up with the minima of the other at
$t+\pi/4\,\om$, to produce zero contrast. This results in an
oscillatory growth and decay of the contrast of the overall
pattern with time. The recoil induced phase can be determined from
this temporally oscillating contrast.

The time evolution of this contrast can be monitored by
continuously reflecting a weak probe beam from the grating (the
additional grating formed by paths 1 and 3 has period $\lambda/4$,
which does not reflect the probe beam). The reflected signal can
be written as
\begin{eqnarray}
\label{signal}
S(T,t)&=&C(T,t)\,\sin^2\left(\frac{\Phi_1(t)+\Phi_3(t)}{2}-\Phi_2(t)\right)\nonumber\\
&=&C(T,t)\,\sin^2(8\,\om T + 4\,\om (t-2T)),
\end{eqnarray}
where $C(T,t)$ is an envelope function whose width is the grating
coherence time, $\frac{1}{k\Delta v}$. This motivated our use of a
BEC atom source. This allowed many contrast oscillations in a
single shot. Using the phase of the reflection at $t\!=\!2T$,
$\Phi(T)=8\,\om T$, $\om$ can be determined by varying $T$.
Vibrational phase shifts and the effect of magnetic bias fields
gradients cancel in the evaluation of
$\frac{\Phi_1(t)+\Phi_3(t)}{2}-\Phi_2(t)$, due to the symmetry of
our scheme.

In this experiment, we realize the scheme of
Fig.$\,$\ref{fig:scheme}(b) and measure $\omega_{\rm rec,Na}$ to
$7\,$ppm precision. We also demonstrate the insensitivity of the
contrast signal to vibrations and the $N^2$ scaling of the recoil
phase.
%--- end of introduction of our interferometer-----------------------

%--- experimental setup ---------------------------------------------
We used sodium BEC's containing a few million atoms in the
$|F\!=\!1,m_F\!=\!-1\rangle$ state as our atom source. The light
pulses were applied $\approx\!15\,$ms after releasing the BEC from
a weak magnetic trap. This lowered the peak density to about
$10^{13}\,$cm$^{-3}$, thus preventing superradiance effects
\cite{inou99} and reducing frequency shifts from mean field
interactions. Two horizontal counterpropagating (to
$\leq\!1\,$mrad) laser beams were used for the diffraction
gratings. The light for the gratings was red-detuned by $1.8\,$GHz
from the sodium $D_2$ line. Rapid switching ($<\!100\,$ns) and
intensity control of the light pulses was done by an acousto-optic
modulator (AOM) common to the two beams. The phase and frequency
of each beam were controlled by two additional AOMs, driven by two
phase-locked frequency synthesizers.

The \ifm pulse sequence was started with a $1\,\mu$s, square
Kapitza-Dirac pulse, centered at $t\!=\!0$. We adjusted the beam
intensity, to put $\approx\!25\,\%$ of the condensate in each of
the $|\pm2\hbar k \rangle$ diffracted orders. This choice yielded
the best final contrast signal. The second order Bragg pulse was
centered at $t\!=\!T$ and was close to Gaussian shaped with a
width of $7.6\,\mu$s. The intensity was chosen to effect a
$\pi$-pulse between the $|\pm2\hbar k \rangle$ states. The smooth
pulse shape reduced the off-resonant population of undesired
momentum states, yielding a transfer efficiency of $>\!90\%$. The
third pulse, used for reading out the contrast signal, was
centered at $t\!=\!2T$ and was typically $50\,\mu$s long. One of
the Bragg beams was used as the readout beam while the other was
blocked.
%--- end of experimental setup ---------------------------------------

%--- the results: interferometer signal ------------------------------

The light reflected from the atoms was separated from the readout
beam path using a beamsplitter and directed by an imaging lens
onto a photomultiplier. A typical \ifm signal is shown in
Fig.$\,$\ref{fig:signal}. We observed the expected contrast
oscillations at frequency $8\,\om$, corresponding to a $5\,\mu$s
period for sodium. We obtained the recoil phase $\Phi(T)$ from the
contrast signal by fitting to a sinusoidal function as in
Eq.$\,$\ref{signal}.

\begin{figure}[htbf]
\begin{center}
\vskip-4mm \epsfxsize=80mm {\epsfbox{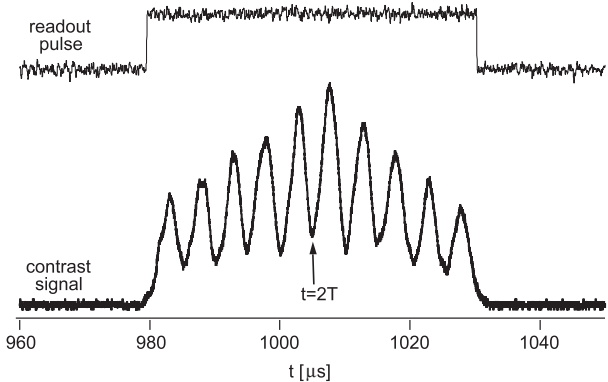}} \vskip-5mm
\end{center}
\caption{Typical single shot signal from the contrast
interferometer. $T\!=\!502.5\,\mu$s, for this example. Ten
oscillations with $\approx\!60\,\%$ contrast and
$\approx\!30\,\mu$s width are observed during the $50\,\mu$s
readout. A low-pass filter at $300\,$kHz (12dB per octave) was
applied to the signal.} \label{fig:signal}
\end{figure}
\vskip-3mm

The signal also contained a small pedestal of similar width as the
envelope. This consists of a constant offset from residual
background light and a smoothly varying contribution from a small
asymmetry between the $|\pm2\hbar k \rangle$ amplitudes of
$<\!5\%$. This asymmetry creates a non oscillating component of
the $2k$ matter wave grating which decays with the same coherence
time. The uncertainty of the fitted phase is about $10\,$mrad,
even if we neglect the envelope function, and assume a constant
amplitude extended over a few central fringes. Similar uncertainty
was obtained for large times $T\!\approx\!3\,$ms. We observe a
shot-to-shot variation in the fitted value of the phase of about
$200\,$mrad. We attribute this to pulse intensity fluctuations
which randomly populated undesired momentum states at the
$<\!10\%$ level. This resulted in spurious matter wave gratings
which shifted the observed recoil phase.
%--- end of: the results, interferometer signal ---------------------

%--- results: recoil measurement ------------------------------------
\begin{figure}[htbf]
\begin{center}
\vskip1mm \epsfxsize=80mm {\epsfbox{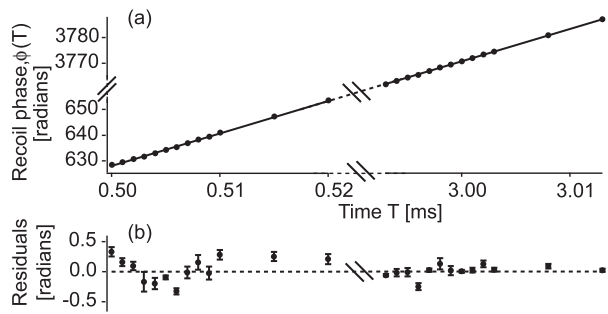}} \vskip4mm
\end{center}
\vskip-5mm \caption{Measurement of $\om$ in sodium. Two sets of
recoil phase scans, around $T\!=\!0.5\,$ms and $T\!=\!3\,$ms, are
shown in (a). Each point is the average of five measurements. The
slope of the linear fit gave $\om$ to $7\,$ppm. The error bars
($\approx 0.05-0.1\,$rads) are shown with the fit residuals in
(b).} \label{fig:recoilmeas}
\end{figure}

\vskip-3mm

The recoil frequency was determined by measuring the recoil phase
around $T\!=\!0.5\,$ms and around $T\!=\!3\,$ms
(Fig.$\,$\ref{fig:recoilmeas}). An upper bound on $T$ was set by
the atoms falling out of the $2\,$mm diameter beam. A straight
line fit to these data produced a value for the sodium photon
recoil frequency $\omega_{\rm rec,Na} = 2\pi \times
24.9973\,$kHz$(1\pm6.7\times 10^{-6})$. This is $2\times 10^{-4}$
lower than the sub-ppm value calculated using the published
measurements of $\alpha_{g-2}$, $R_\infty$, $M_{\rm Na}$
\cite{brad99}, $M_e$, and $\lambda_{\rm Na} \cite{junc81}$ in
Eqs.$\,$\ref{alpha} and $\,$\ref{omega}. The systematic mean field
shift due to larger population in the middle path than the extreme
paths probably explains this deviation. Estimated errors from beam
misalignment and wavefront curvature have the same sign as the
observed deviation but several times lower magnitude.
%--- end of: results, recoil measurement -----------------------------

%--- results: phase insensitivity ------------------------------------
\begin{figure}[htbf]
\begin{center}
\vskip-1mm \epsfxsize=80mm {\epsfbox{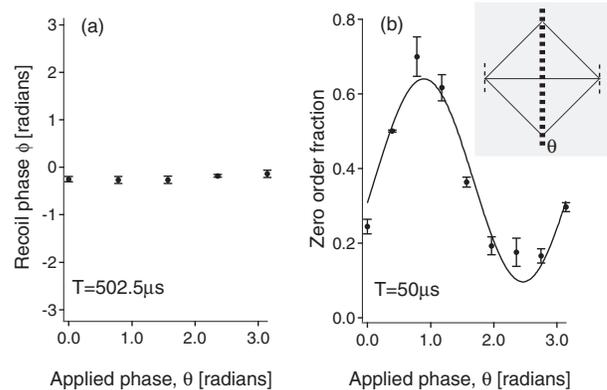}} \vskip0mm
\end{center}
\caption{Vibration insensitivity of the contrast interferometer.
(a) shows the measured recoil phase at $T\!=\!502.5\,\mu$s from
the contrast \ifm as a function of the applied phase $\theta$. The
recoil phase is constant and demonstrates our insensitivity to
phase noise from the gratings. (b) shows the fractional population
of the \stz state from the phase-sensitive \ifm (inset) for a
similar scan of $\theta$ at $T\!=\!50\,\mu$s. Also shown is the
best-fit sinusoid of the expected period.} \label{fig:phasestab}
\end{figure}

\vskip-3mm

To demonstrate the insensitivity of the measurement to phase noise
of the light due to mirror vibrations, we intentionally varied the
phase $\theta$ of the second grating relative to the first one
\cite{foot2}. The contrast signal is not visibly affected by such
phase variations (Fig.$\,$\ref{fig:phasestab}(a)). We compared
this to a phase-sensitive readout method
(Fig.$\,$\ref{fig:phasestab}(b), inset). This was realized by
replacing the readout pulse with a third pulsed $1\,\mu$s light
grating in the Kapitza-Dirac regime, phase-locked to the first two
pulses. This projected the phase of the $2k$ pattern at $t\!=\!2T$
onto the fractional populations of the states $|0\,\hbar
k\rangle$, $|2\,\hbar k \rangle$, and \stmt which leave this
interferometer. The populations were measured by time-of-flight
absorption imaging. The \stz fraction is shown for the same
variation of $\theta$, in Fig.$\,$\ref{fig:phasestab}(b). The
oscillation \cite{foot3} demonstrates the phase sensitivity of any
position-sensitive readout.

These two interferometers respond differently to mirror
vibrations. For large $T$, we have observed the effect of the
mirror vibrations directly. At $T\!\approx\!3\,$ms, the
shot-to-shot fluctuations of the phase-sensitive \ifm was of the
order of the expected fringe contrast. This agrees with
observations with a standard Mach-Zehnder \ifm constructed both by
us and in \cite{tori00}. In comparison, the stability of the
contrast \ifm signal is independent of $T$ within our
measurements. This can be seen from the comparable statistical
error bars for short and long times in
Fig.$\,$\ref{fig:recoilmeas}(b). In fact, the residuals and the
corresponding error bars are smaller at the longer times. We
attribute this to the decreased amplitude of some of the spurious
gratings at longer times, due to reduced overlap of the
contributing wavepackets.
%--- end of: results, phase insensitivity ---------------------------

%--- results: demonstrating the N^2 effect --------------------------
\begin{figure}[htbf]
\begin{center}
\vskip-1mm \epsfxsize=80mm \hspace{0cm}{\epsfbox{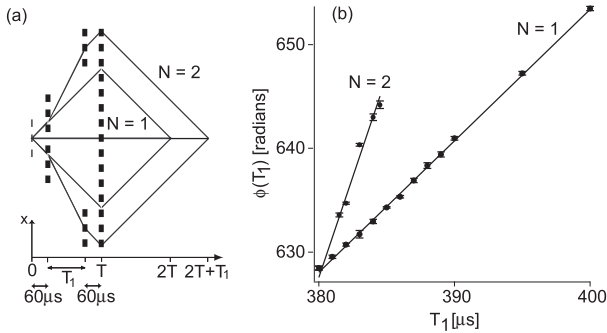}}
\vskip2mm
\end{center}
\vskip-4mm \caption{Demonstration of the quadratic scaling of the
recoil phase with additional photon recoils. (a) shows the
$N\!=\!1$ (inner) and $N\!=\!2$ (outer) interferometers used. (b)
shows the recoil phase at the recombination time under variation
of $T_1$.} \label{fig:Nsquare}
\end{figure}

\vskip-3mm

The quadratic scaling of the accumulated recoil phase with the
number of transferred recoils $N$, was demonstrated by comparing
$N\!=\!1$ and $N\!=\!2$ interferometers. An $N\!=\!2$ geometry,
shown in Fig.$\,$\ref{fig:Nsquare}(a), was realized using two
additional first order Bragg $\pi$-pulses spaced $T_1$ apart and
affecting only the extreme paths. The acceleration pulse at
$t\!=\!60\,\mu$s drove transfers from $|\pm2\hbar k \rangle$ to
$|\pm4\hbar k \rangle$. The deceleration pulse at
$t\!=\!T_1+60\,\mu$s$\,=\!T-60\,\mu$s drove transfers from
$|\pm4\hbar k \rangle$ to $|\pm2\hbar k \rangle$. During the
period $T_1$, paths 1 and 3 accumulate phase $2^2=4$ times faster
in the $N\!=\!2$ scheme than in the $N\!=\!1$ scheme. Additional
time $T_1$ is required for the three paths to overlap in the
$N\!=\!2$ scheme. For this geometry, the $N\!=\!2$ recoil phase
should therefore evolve three times faster as a function of $T_1$
than the $N\!=\!1$ recoil phase. The corresponding data sets are
shown in Fig.$\,$\ref{fig:Nsquare}(b). The linear fits give a
slope ratio of $3.06\pm0.1$. At present, we do not have sufficient
control over the timing and phase of the intermediate pulses to
improve our $N\!=\!1$ measurement precision by using $N\!>\!1$.
%--- end: results, demonstrating the N^2 effect ---------------------

%--- discussion & conclusions ---------------------------------------
In conclusion, we have demonstrated a contrast \ifm which has
several desirable features for a high precision measurement of the
photon recoil frequency. Such a measurement would involve
converting to an atomic fountain setup with vertical Bragg beams.
In this geometry, $T$ can be extended by nearly two orders of
magnitude. Our insensitivity to phase noise from mirror vibrations
should greatly alleviate vibration isolation requirements of the
system for long $T$ \cite{youn97}. The order $N$ of the \ifm must
also be increased, requiring improved timing and phase control of
laser pulses. Direct scaling of our current precision of $\approx
0.01\,$rads/shot, results in an estimated precision of
$<1\,$ppb/shot for $T\!=\!100\,$ms and $N\!=\!20$. A rigorous
study of systematics will have to be undertaken to increase the
accuracy of our measurement. Estimates show that mean field
effects can be suppressed to the ppb level by reduction of atomic
density to $\approx 10^{11}\,$cm$^{-3}$, together with pulse
control for $<5\%$ imbalance between populations in the middle and
extreme paths. In addition, our methods may provide a way to study
mean field effects with interferometric precision. We hope to
obtain a $<\!1\,$ppb value for $\om$ in a second generation
experiment in which BECs are created elsewhere and transported
into the \ifm\cite{gust02}.
%--- end: discussion & conclusions ---------------------

We thank W. Ketterle for valuable discussions. This work was
supported by the NSF, ONR, ARO, NASA and the David and Lucile
Packard Foundation.


\vspace{-0.7cm}
\begin{thebibliography}{10}
\vspace{-0.8cm}

\bibitem{dyck87}
R.~S. Van~Dyck, Jr., P.~B. Schwinberg, and H.~G. Dehmelt, Phys.
Rev. Lett. {\bf 59}, 26 (1987).

\bibitem{kino95}
T. Kinoshita, Phys. Rev. Lett. {\bf 75}, 4728 (1995).

\bibitem{cage89}
M.~E. Cage et. al., IEEE Trans. Instr. Meas. {\bf 38}, 284 (1989).

\bibitem{tayl94}
B. Taylor, Metrologia {\bf 31}, 181 (1994).

\bibitem{schw99}
C. Schwob et. al., Phys. Rev. Lett. {\bf 82}, 4960 (1999).

\bibitem{udem97}
Th. Udem et. al., Phys. Rev. Lett. {\bf 79}, 2646 (1997).

\bibitem{beie02}
T. Beier et. al., Phys. Rev. Lett. {\bf 88}, 011603 (2002).

\bibitem{krug98}
E. Krüger, W. Nistler, and W. Weirauch, Metrologia {\bf 35}, 203
(1998).

\bibitem{brad99}
M.~P. Bradley et. al., Phys. Rev. Lett. {\bf 83}, 4510 (1999).

\bibitem{udem99}
Th. Udem et. al., Phys. Rev. Lett. {\bf 82}, 3568 (1999).

\bibitem{foot1}
$6\,$ppb \cite{hens01};$15\,$ppb (group website).

\bibitem{weis93}
D.~S. Weiss, B.~C. Young, and S. Chu, Phys. Rev. Lett. {\bf 70},
2706 (1993).

\bibitem{youn97}
B.~C. Young, Ph.D. thesis, Stanford (1997).

\bibitem{hens01}
J.~M. Hensley, Ph.D. thesis, Stanford (2001).

\bibitem{cahn97}
S.~B. Cahn et. al., Phys. Rev. Lett. {\bf 79}, 784 (1997).

\bibitem{berm97}
See pp. 281-282 and 379-381 in {\it Atom Interferomtery}, edited
by P. Berman (Academic Press, New York, 1997).

\bibitem{weit97}
M. Weitz, T. Heupel, and T.~W. Hänsch, Appl. Phys. B. {\bf 65},
713  (1997).

\bibitem{gupt01}
S. Gupta et. al., Cr. Acad. Sci. IV-Phys {\bf 2}, 479 (2001), and
references therein.

\bibitem{inou99}
S. Inouye et. al., Science {\bf 285},  571  (1999).

\bibitem{junc81}
P. Juncar et. al., Metrologia {\bf 17},  77 (1981).

\bibitem{foot2}
We scanned $\theta$ by electronically shifting the phase of the rf
signal used to drive one of the two Bragg AOMs  .

\bibitem{foot3}
The division of population into {\it three} output ports caused
$74\%$ ($<\!100\%$) contrast. We have seen $\approx\!100\%$
contrast in a standard Mach-Zehnder interferometer.

\bibitem{tori00}
Y. Torii et. al., Phys. Rev. A. {\bf 61},  041602  (2000).

\bibitem{gust02}
T.~L. Gustavson et. al., Phys. Rev. Lett. {\bf 88},  020401
(2002).

\end{thebibliography}
\end{document}